\documentclass[10pt,conference]{IEEEtran}
\usepackage{amsmath,amssymb,amsfonts}
\usepackage{algorithmic}
\usepackage{graphicx}
\usepackage{textcomp}
\usepackage{url}
\usepackage[natbib=true, sorting=none, firstinits=true, maxcitenames=1, style=numeric-comp]{biblatex}

\def\BibTeX{{\rm B\kern-.05em{\sc i\kern-.025em b}\kern-.08em
    T\kern-.1667em\lower.7ex\hbox{E}\kern-.125emX}}

\newlength{\xfigwd}
\title{Systematic benchmarking of quantum computers: status and recommendations}

\author{Jeanette Miriam Lorenz\IEEEauthorrefmark{1}\IEEEauthorrefmark{2},
Thomas Monz\IEEEauthorrefmark{3}\IEEEauthorrefmark{4},
Jens Eisert\IEEEauthorrefmark{5}\IEEEauthorrefmark{6},
Daniel Reitzner\IEEEauthorrefmark{7},
Félicien Schopfer\IEEEauthorrefmark{8},\\
Frédéric Barbaresco\IEEEauthorrefmark{9}, 
Krzysztof Kurowski\IEEEauthorrefmark{10},
Ward van der Schoot\IEEEauthorrefmark{11},
Thomas Strohm\IEEEauthorrefmark{12},
Jean Senellart\IEEEauthorrefmark{13},\\
Cécile M.\ Perrault\IEEEauthorrefmark{14},
Martin Knufinke\IEEEauthorrefmark{15},
Ziyad Amodjee\IEEEauthorrefmark{16},
Mattia Giardini\IEEEauthorrefmark{17}\\
\IEEEauthorblockA{\IEEEauthorrefmark{1}Fraunhofer Institute for Cognitive Systems IKS, 80686 Munich, Germany (email: jeanette.miriam.lorenz@iks.fraunhofer.de)}
\IEEEauthorblockA{\IEEEauthorrefmark{2}Ludwig-Maximilians-University Munich, 80539 Munich, Germany}
\IEEEauthorblockA{\IEEEauthorrefmark{3}Universität Innsbruck, Institut für Experimentalphysik, Technikerstrasse 25, 6020 Innsbruck, Austria}
\IEEEauthorblockA{\IEEEauthorrefmark{4}Alpine Quantum Technologies GmbH, 6020 Innsbruck, Austria}
\IEEEauthorblockA{\IEEEauthorrefmark{5}Dahlem Center for Complex Quantum Systems, Freie Universität Berlin, 14195 Berlin, Germany}
\IEEEauthorblockA{\IEEEauthorrefmark{6}Fraunhofer Heinrich Hertz Institute, 10587 Berlin, Germany}
\IEEEauthorblockA{\IEEEauthorrefmark{7}CyberSecurity Hub, z. ú., Šumavská 416/15, 602 00 Brno, Czech Republic}
\IEEEauthorblockA{\IEEEauthorrefmark{8}Laboratoire national de métrologie et d'essais, 29 avenue Roger Hennequin, 78190 Trappes, France}
\IEEEauthorblockA{\IEEEauthorrefmark{9}THALES, 1 Av. Augustin Fresnel, 91120 Palaiseau, France}
\IEEEauthorblockA{\IEEEauthorrefmark{10}Poznan Supercomputing and Networking Center IBCH PAS, Noskowskiego 12/14, 61-704 Poznan, Poland}
\IEEEauthorblockA{\IEEEauthorrefmark{11}The Netherlands Organisation (TNO) for applied scientific research, The Hague}
\IEEEauthorblockA{\IEEEauthorrefmark{12}Robert Bosch GmbH, Corporate Research, D-71272 Renningen, Germany}
\IEEEauthorblockA{\IEEEauthorrefmark{13}Quandela, 7 rue Léonard de Vinci, 91400 Massy, France}
\IEEEauthorblockA{\IEEEauthorrefmark{14}Alice $\&$ Bob, 49 boulevard du general Martial Valin, 75015 Paris, France}
\IEEEauthorblockA{\IEEEauthorrefmark{15}Eviden, science+computing AG, Hagellocher Weg 73, 72070 Tübingen, Germany}
\IEEEauthorblockA{\IEEEauthorrefmark{16}Laboratoire Charles Fabry, Institut d’Optique Graduate School, CNRS, Universite Paris Saclay, 91127 Palaiseau, France}
\IEEEauthorblockA{\IEEEauthorrefmark{17}European Quantum Industry Consortium e.V.,Leo-Brandt-Straße, 52428 Jülich }
}

\begin{document}

\maketitle

\begin{abstract}
Architectures for quantum computing can only be scaled up when they are accompanied by suitable benchmarking techniques. The document provides a comprehensive overview of the state and recommendations for systematic benchmarking of quantum computers. Benchmarking is crucial for assessing the performance of quantum computers, including the hardware, software, as well as algorithms and applications. The document highlights key aspects such as component-level, system-level, software-level, HPC-level, and application-level benchmarks. Component-level benchmarks focus on the performance of individual qubits and gates, while system-level benchmarks evaluate the entire quantum processor. Software-level benchmarks consider the compiler's efficiency and error mitigation techniques. HPC-level and cloud benchmarks address integration with classical systems and cloud platforms, respectively. Application-level benchmarks measure performance in real-world use cases. The document also discusses the importance of standardization to ensure reproducibility and comparability of benchmarks, and highlights ongoing efforts in the quantum computing community towards establishing these benchmarks. Recommendations for future steps emphasize the need for developing standardized evaluation routines and integrating benchmarks with broader quantum technology activities.
\end{abstract}


\tableofcontents

\section*{List of abbreviations}

\begin{description}
\item[ACES] \textit{Averaged circuit eigenvalue sampling} is a component-level benchmarking method that is capable of detecting and characterising correlations.

\item[FTQC] \textit{Fault-tolerant quantum computing} relies on careful computational instructions such that errors do not
spread during QC in an uncontrollable fashion. These carefully designed instructions are
accompanied by QEC routines that remove undesired errors during the computation and allow, 
in principle, for arbitrarily long reliable computations.

\item[KPI] \textit{Key performance indicators} are measured according to some suitable metrics and are derived from several benchmarking methods.

\item[MCDA] \textit{Multi-criteria decision aiding} methods are techniques designed to evaluate multiple, often conflicting criteria to aid decision-making and ranking.

\item[NISQ] \textit{Noisy intermediate-scale quantum} \cite{PreskillNISQ}
devices describe quantum devices available at present and in the near future that do not allow FTQC.
They contain essentially all quantum and hybrid computational schemes that are not contained within the group of FTQC.

\item[QC] \textit{Quantum computing}.

\item[QEC] \textit{Quantum error correction} embodies schemes for error reduction in QC that rely on quantum redundant encoding and measurements which need not be necessarily fault tolerant.

\item[RB] \textit{Randomized benchmarking} refers to a family of benchmarking processes in which a set of gate operations is randomly generated such that the total sequence ideally corresponds to the identity operation. The decay of the success rate as a function of the length of the sequence allows one to infer the average error rate in quantum gate sets.

\item[SDK] \textit{Software Development Kit}.

\end{description}

\section{Introduction}
\label{sec:introduction}

While quantum computers were already proposed in the 1980s, their practical usefulness remains an open question despite substantial recent breakthroughs in realizing fault-tolerant quantum devices
in the last decade~\cite{GoogleQuantumAIandCollaborators:2024efv,Evered:2023wbk}. Potential applications of quantum computing include all academic or industrial use cases which rely on simulating quantum mechanical systems (such as, e.g., in drug development or materials science), on solving linear systems of equations, on addressing combinatorial optimization challenges and on enhancing machine learning by suitable quantum subroutines.

The computational paradigm of quantum computing was first proposed by 
D.\ Deutsch in 1985, 
who thought about a stronger version of the Church-Turing thesis. Richard Feynman pointed out in his 1982 article~\cite{feynman1982simulating} that classical computers may have difficulties in simulating quantum systems and that quantum devices may be of use here, and added substantial insights into notions of mutual simulatability and computational efforts. Only in the mid 1990s, 
a key quantum algorithms was proposed by Peter Shor~\cite{shor1994}, providing a practically relevant algorithm to factor big numbers into their prime factors, featuring a super-polynomial advantage over the best known classical algorithms. Lov K.\ Grover around the same time proposed an algorithm for an efficient search through unsorted databases \cite{Grover1996}, featuring a polynomial quantum advantage.

First experimental realizations of small noisy quantum computers were established in laboratory settings in the 1990s, but they could not yet be used to run any extended computations. This state of affairs started to change only more than 20 years later: In 2019, Google demonstrated an experiment on the Google Sycamore chip featuring 53 superconducting qubits. In this experiment, Google claimed
a quantum advantage by solving a specific mathematical problem significantly faster than what was possible on any available classical supercomputer~\cite{Arute:2019zxq}. Although this claim was 
challenged later by classical simulations \cite{PhysRevX.10.041038,SupremacyReview}  --   but also subsequently
strengthened by more sophisticated experiments \cite{PanSampling,GoogleQuantumAIandCollaborators:2024efv,PhysRevLett.134.090601}  --   
it did trigger an increased interest in quantum computing as a technology and in its disruptive potential for a significant number of industrial challenges. Similarly, in 2020, the group of J.-W. Pan claimed a quantum advantage on Gaussian boson sampling~\cite{PhysRevLett.123.250503}, a
specific quantum random sampling scheme based on photonic Gaussian input states.
In 2023, a team of IBM
suggested having reached the regime of quantum utility~\cite{Kim:2023bwr}, which constitutes a regime where quantum computers are comparable to classical computers. The state of the art chip by Google Quantum AI is the 
Willow chip, a  72-qubit and a
 105-qubit quantum processor designed for quantum error 
 correction experiments
 \cite{GoogleQuantumAIandCollaborators:2024efv}.

By now, various quantum hardware vendors have started to offer access to their quantum computers for academic and industrial use, partly via cloud access. This has triggered work towards practical testing of quantum computer capabilities. However, despite various claims of quantum supremacy, advantage or utility, usefulness of quantum computers for concrete tasks that tackle immediate practical industrially relevant applications
have not be demonstrated yet. Significant advances on the quantum hardware, software and quantum algorithm side are required before this could be demonstrated. For instance, different works~\cite{hoefler2023disentangling,Santagati:2023lkf} 
have shown that currently proposed algorithms such as, e.g., Grover’s algorithm, would result in longer execution times than those of classical algorithms even if fault-tolerant quantum computers would exist. This is because typical quantum gate operation times on quantum computers (in the $ns$ to $\mu s$ regimes) are typically longer than gate operation times on classical computers. Therefore, quantum algorithms may need to provide an exponential or super-polynomial computational speed-up in comparison to classical computers~\cite{hoefler2023disentangling} to practically accelerate calculations time-wise.

Additionally, the significant progress over the last years in the area of quantum hardware and software development have led to first attempts at quantum software stacks~\cite{software_stack}. The goal of 
such stacks
is to implement abstraction layers for using quantum computers similar to what is employed in classical computers. These software development activities are intended to make quantum computers more broadly available for academics, industry, and the general public. One goal of these software developments aims at extending from purely quantum algorithms to hybrid quantum-classical algorithms, which will be composed of both quantum and classical computing infrastructure. As a consequence, to assess whether the use of a quantum computer is beneficial to solve a given use case, one will need to assess both the performance of the quantum part of a computation, as well as the performance of all other elements in the software stack, including the classical part of the hybrid algorithm.

Given this exciting development, during the recent years, more and more questions have arisen on how the performance of quantum computers, algorithms and software can be assessed and compared; consequently, various \textbf{benchmarks} have been proposed.

Moving to the core topic of this article, in this development, benchmarks are crucially tools in assessing the performance of implementations of quantum computers. 
They are generally associated with figures of merit or \textbf{key performance indicators} (KPI) that are measured according to some suitable protocol (see Fig.\ 1). Benchmarks are typically classified into three levels:

\begin{itemize}
\item 
\textbf{Component-level}: The first such level concerns the hardware implementation of individual components~\cite{eisert2020quantum}. One well-established figure of merit is the average gate fidelity that captures the quality of a gate-set in quantum computing and which is commonly estimated by means of randomized benchmarking techniques~\cite{Emerson+05,KnillBenchmarking,Dankert+09,MagGamEmer}. Other similar quantities would be channel fidelities. There are a multitude of such figures of merit. Usually, there are trade-offs in the assumptions one is willing to make and the sample complexities of the scheme, as well as in the information gain that is anticipated about the performance of the component and the number of samples needed.
\item
\textbf{System-level}: The second level captures the hardware performance of entire implementations. The quantum volume~\cite{Cross+19} 
is such a figure of merit, quantifying the capabilities and error rates of an entire quantum circuit composed of quantum gates. It captures the maximum size of square quantum circuits that can be implemented reliably by the quantum circuit. The layer fidelity~\cite{LayerFidelity} 
is a similar such quantity that provides a benchmark that measures the entire processor’s ability to run circuits while revealing information about the performance of individual qubits and quantum gates.
\item
\textbf{Application-level}: On the highest level, one often compares the performance of entire families of protocols.
\end{itemize}

We have also identified additional finer structure to benchmarking quantum computers that includes software level, HPC and cloud level, and various classical metrics, though these levels are not yet as well developed in the quantum context and community.

Different European initiatives have been started during recent years to develop certain benchmarks, such as, e.g., the French BACQ~\cite{Barbaresco:2024fmg} and German Bench-QC~\cite{Bench-QC} with the intent to develop benchmarks that are presumably useful to assess the overall performance of quantum computers for various use cases. However, given the multitude of initiatives, currently an enormous amount of benchmarks on different levels and with different purposes had been proposed, without having established a standard or a generally agreed benchmarking protocol. 
Motivated by this state of affairs, this paper intends to summarize and discuss various available benchmarks to establish a consensus between the involved research groups in which direction the development and adoption of benchmarks should go in the future.

\section{Terminology}
\label{sec:terminology}

There are numerous quantum paradigms developed to complement existing classical computing capabilities. While these approaches may significantly differ on a technological level,
they can be broadly categorized based on their control methodology as follows.

\begin{itemize}
\item \textbf{Digital universal quantum computers} are built on a quantum register (addressable set of qubits) that is initialised, manipulated by a discrete set of gate operations acting on specific qubits, and subsequently read out. The discrete nature allows the computation to be interrupted and resumed throughout the computation, without a change to the result. Interruption at mid-execution allows for error mitigation and error correction methods to be applied, to potentially improve system performance. Throughout the computation, the system is typically meant to maintain a high level of coherence. Universality refers to the feature that an arbitrary unitary gate operation can be arbitrarily well approximated.

\item \textbf{Analog quantum computers} consist of a quantum constituents that are initialised, manipulated by continuous interactions of variable strength, and read out at the end. Given the continuous interactions, 
basically by Hamiltonian control, interruption during the computation is not easily possible without changing the result. Throughout the computation, the system is typically meant to maintain high coherence. 
Some programmable quantum architectures can be seen as being intermediate between digital and analog quantum computers.

\item \textbf{Quantum annealers} are similar in nature to analog quantum computers, yet, employ incoherent or coherent effects such as tunnelling
during the computation. This does not negatively impact the results, but affects the computations and applications that can be implemented in such systems.

\item \textbf{Quantum random sampling schemes} such as Gaussian boson
or universal circuit sampling can be seen as a programmable ``quantum Galton board''. Such schemes have been set up to design tasks for which there is strong evidence that existing quantum hardware platforms can outperform classical computers for specific tasks and hence achieve a quantum advantage. This functionality affects the range of applications that can be implemented. Photonic architectures for boson sampling that include measurements can be uplifted to fully universal quantum computational schemes.
\end{itemize}

In the context of this document on notions of benchmarking, we shall restrict ourselves to benchmarks suitable for digital universal
quantum computers. Within this field and due to its discrete nature, building blocks are comparably clearly defined, which can all be characterised either component-wise or holistically. The possibility of error correction provides the confidence in scalability of these technologies. Unlike the other three approaches, digital quantum computing has well-established quantum software development kits that support the realisation of standardized benchmarks that can and shall be implemented in a hardware-agnostic fashion.

Nevertheless, some, and, in particular, most application-oriented benchmarks can be applied not only to digital quantum computing, but also to the three other approaches in order to assess their capabilities and performance on concrete use cases on the same footing.

\section{State of the art of quantum computing benchmarking world-wide}
\label{sec:SOTA}

Benchmarks have historically evolved in the field of quantum information sciences along the maturation of quantum information processing. Quantum benchmarking, therefore, started by covering \textbf{component analysis}, such as the fidelity of states or processes realised. These approaches attempt at a full characterization of the considered component and are, therefore, typically hard to scale.
To our knowledge, the largest quantum state that has been fully characterized through quantum state tomography so far consists of 8 qubits~\cite{haffner2005scalable}, while the largest unknown process 
that has been fully characterized has
involved~3 qubits~\cite{PhysRevLett.102.040501}. For comparison, some technology platforms offer about~1000 qubits to date \cite{Condor}.
The main challenge here is that it is generally not feasible to extrapolate performance of component characterisations of a few qubits to hundreds, if not thousands of qubits. Due to the currently perceived lack of being able to predict system performance from component analysis motivates why methods offering limited value for a potential users will not be further discussed in the context of this document.

The overall system performance is dominated by the performance of these operations that take place most often, or have exceedingly bad performance. For NISQ computing, with a quantum computation consisting of initialisation, manipulation, and readout, the number of gate operations typically significantly exceeds the number of initialisation or readout operations. Therefore, the gate performance mainly determines the overall system performance. In the case of FTQC, the additional interleaved QEC steps are still dominated by manipulation (e.g., syndrome mapping) compared to qubit readout and qubit reset.
The efficient characterisation of gate operations is the subject of significant effort and resulted in novel methods that would not describe each and every aspect of a gate operation (with its exponentially growing number of parameters), but provide a single number: the probability to implement the desired operation, or alternatively: the error rate for an intended process. Here, the class of \textbf{randomized benchmarking} (RB) protocols~\cite{Emerson+05,KnillBenchmarking,Dankert+09,MagGamEmer} covers a wide range of tools to describe the success probability for gate operations in a scalable fashion.

With a growing number of qubits and gate operations available, people have started to combine these computational building blocks. These first realisations of more complex experiments demonstrated that sometimes the combination of components behaves differently from what was expected. Effects such as cross-talk, heating, and other platform-dependent detrimental influences showed that there may be temporal and spatial correlations, non-Markovian processes, and other effects that reduce the predictability of the system behavior. The research community therefore started to develop holistic benchmarks to describe the performance of large systems, ideally using protocols that are scalable, to provide a statement about the overall performance of a device. Prominent examples include (logarithmic) quantum volume, or the \textit{cross-entropy benchmark} (XEB)~\cite{Boixo+18,SupremacyReview} or variants thereof that are
still being used as state-of-the-art benchmarks of quantum
processors \cite{GoogleQuantumAIandCollaborators:2024efv}.
Besides the performance, recent benchmarks also try to cover aspects such as time-to-solution, for instance by counting operations per second, specifically \textbf{circuit layer operations per second} (CLOPS)
\cite{LayerFidelity,Wack:2021gvg}, or energy consumed.

These efforts are also intertwined with efforts within 
\textit{high-performance computing} (HPC). Europe is currently leading the integration of quantum computers into HPC facilities. The EuroHPC Joint Undertaking\footnote{\url{https://eurohpc-ju.europa.eu/index_en}} has launched an initiative to acquire up to eight quantum computers, which will be integrated with classical supercomputers in European HPC facilities. To assess the overall performance of each EuroHPC quantum computer before the final selection process, significant efforts have been made to develop comprehensive benchmarks that incorporate both hardware and application metrics~\cite{EuroQCS}. However, these evaluation processes have focused primarily on different quantum technology platforms rather than on their performance in fully integrated classical-quantum setups, due to the lack of available reference installations.

With quantum computers being attached to an HPC facility as an instance of what could be called a quantum accelerator, several questions arise for which there are currently no existing benchmarks: What level of integration do we specifically talk about? Can the HPC system access the coherent memory in real-time? What is the latency for communication between the HPC system and the quantum processor in each direction? How efficient is the compilation of instructions from the HPC system onto the QC system? There are currently no libraries, even less standardised routines, to investigate questions in this domain.

End users may not care about aspects such as latencies, fidelities of individual gate operations, or how the quantum computer is integrated within the classical infrastructure, but may focus on \textbf{applications}: is the answer correct? How long will the computation take? Can I improve the accuracy of the calculation 
(e.g., with error mitigation)? If so, by how much? These questions will subsequently result in the question: there are different quantum technology platforms. Which platform is most suited for which applications? As long as quantum computers work with relatively small number of qubits (on the order of 100 or less), these questions and their answers can be cross-checked using classical computers. But, what will the community do once the capabilities of quantum processors surpass classical computers? Therefore benchmarks, at an early stage, should be compatible also in the regime beyond classical computational capabilities.

\begin{figure}
\centering\includegraphics[width=\columnwidth]{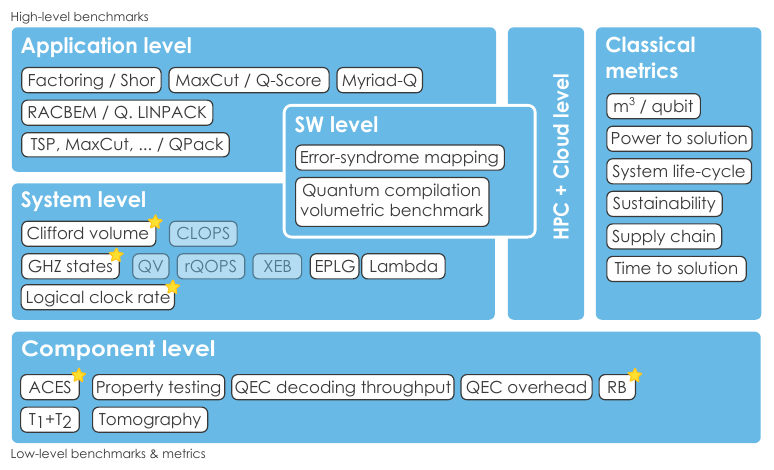}
\caption{Benchmarks at different levels  --  in addition to the typical distinction between a component-level, system-level and application-level, we have identified also the distinction into software-level benchmarks, HPC and cloud-level benchmarks, as well as specific classical metrics. Some benchmarks, especially in the system levels are not scalable. These have been shaded out. Some benchmarks in component or system level, marked by a star, are FTQC-ready and might be useful to compare over time in both regimes, NISQ as well as FTQC. This holds also for all benchmarks in the application level, but for clarity the stars are omitted.}
\label{fig:benchmarks}
\end{figure}

In the following, we list existing initiatives proposing benchmarks internationally:

\begin{itemize}
\item United States of America:
\begin{itemize}
\item \textbf{QED-C Benchmark} by The Quantum Economic Development Consortium~\cite{QED-C},
\item \textbf{QBI: Quantum Benchmark Initiative} by DARPA~\cite{DARPA},
\item \textbf{HAMLIB} by INTEL~\cite{Sawaya:2023nmv},
\item \textbf{MetriQ} by the Unitary Fund~\cite{metriq},
\item \textbf{WG P7131 ``Standard for Quantum Computing Performance Metrics \& Performance Benchmarking"} by IEEE~\cite{IEEE_QCstandards}. 
\end{itemize}
\item United Kingdom:
\begin{itemize}
\item \textbf{UK Quantum Computing Metrics} by UK Quantum Metrology Institute of PNL~\cite{UK_quantum_metrics}.
\end{itemize}
\item Europe
\begin{itemize}
\item \textbf{German Bench-QC project} “Application-driven benchmarking of quantum computers” by Fraunhofer IKS \& IIS, BMW, Optware, ML Reply and Quantinuum~\cite{Bench-QC},
\item \textbf{QUARK:} A Framework for Quantum Computing Application Benchmarking by BMW~\cite{Finzgar:2022mya},
\item \textbf{Dutch QPack project} “Benchmark for quantum computing” by Delft University of Technology~\cite{QPack},
\item \textbf{Dutch TNO quantum metrics Initiative} "Application-level quantum metrics for current and future quantum computers" by TNO~\cite{Mesman+24},
\item \textbf{Spanish CUCO project} “Computación Cuántica En Industrias Estratégicas” by GMV~\cite{CUCO},
\item \textbf{Basque Quantum Benchmark} ``Benchmarking Quantum Computers: Towards a Standard Performance Evaluation Approach"~\cite{Acuaviva:2024pdr},
\item \textbf{French BACQ project} “Benchmarks Applicatifs des Calculateurs Quantiques” by THALES, LNE, CEA, CNRS, EVIDEN \& TERATEC~\cite{Barbaresco:2024fmg},
\item \textbf{Polish Open QBench, EuroQCS-Poland project} "Open QBench: Application Performance Benchmarks for Quantum Computers" by PSNC~\cite{QBench}. 
\end{itemize}
\end{itemize}

\section{How to define quantum benchmarks}
\subsection{Benchmark versus metric}
\label{sec:benchmarc_metric}

In the field of (quantum) benchmarking, the terms benchmark and metric are often used interchangeably. In addition, they are often used to mean different things. For that reason, this section aims to define both of them clearly and fix the terminology. 

A \textbf{(quantum) metric} is a quantitative measure to assess the performance of a (quantum) device, specified by a clear protocol. 
The verb \textbf{(quantum) benchmarking} details the act of performing a (quantum) benchmark.
A \textbf{(quantum) benchmark} is a set of results of implementing one or multiple quantum metric(s) on a certain set of (quantum) devices
aimed at characterizing the accuracy of the implementation.

\subsection{Objectives for benchmarks}
\label{sec:objectives}

Broadly speaking, benchmarks can be used by both specialists as well as the general public. They offer means to turn demonstrations into a quantifiable number, which can be compared across platforms, regions, and time. Ideally the quantified performance also raises awareness for achievements as well as challenges that need to be tackled: from demonstrating the capability to implement large-scale quantum computing via reproducible statements about the accuracy of quantum computing results to the time such computations will take. These benchmarks will help guide researchers and developers towards faster, larger, and better systems, and communicate this progress both within the quantum community as well as to the general public.

While benchmarks are technically motivated to help guide future development and raise awareness for both progress as well as challenges, the numbers, in particular the comparison between devices and regions, will also affect opinions. It is, therefore, important to ensure that benchmarks are implemented and evaluated in a standardised fashion to guarantee both reproducibility as well as comparability. It will, for this reason, be important to establish loophole-free and standardised benchmarks in the long run.

In addition, a good benchmark should be useful also in the future to give a good understanding of the progress and to highlight challenges, therefore, features ensuring its applicability in FTQC should be prefered. One of the most prominent properties that needs to be present in benchmarks already now is their scalability. Many attempts at benchmarking, including quantum volume, suffer from not being scalable, and are, therefore, impossible to be utilized in their current form already in systems of moderate size and well within NISQ era.

Benchmarks are not universal  --  different benchmarks are useful in different contexts. Different stakeholders will have different preferences in what benchmarks should measure. While researchers might be interested in very good system properties, investors might be more interested in overall performance, and industry or end-users might look for best performance towards their goal. All these preferences are valid and give support for different benchmarking levels noted already above.

Finally, there should be hardware-agnostic benchmarks that would compare performance of different platforms. Such comparison will be useful for understanding of overall progress of different platforms. Yet such benchmarks will hardly offer indications on what QC architecture fits best specific use cases. For this reason there is also a need for hardware-dependent benchmarks.

The wide variety of demands on benchmarks makes it even more necessary to have a more rigid (standardized) way of defining them. The aim of this paper is not to propose benchmarks but rather to provide recommendations on how such a benchmark should look like. This shall benefit both providers and consumers of QC technology. 

\subsection{Procedure to define a benchmark or a protocol}
\label{sec:procedure}

The definition of quantum computing benchmarks requires careful consideration of methodology and implementation details to ensure reproducibility and meaningful comparisons across platforms. Drawing inspiration from classical computing standards such as LINPACK, we establish a comprehensive framework for benchmark specifications. 

Protocols and procedures provide step-by-step descriptions of the benchmarking processes, including the types of circuits used, measurement techniques, and data analysis methods. Flowcharts as diagrams can also outline the benchmarking process, showing the sequence of steps involved in executing a benchmark. Finally, infographics are used as visual summaries that combine text, graphics, and data to present key findings in an accessible format.

Each benchmark protocol shall ideally be defined in an \textit{software development kit} (SDK)-agnostic format to maintain generality and platform independence. While the abstract definition ensures broad applicability, at least one reference implementation using a specific SDK should be provided. This reference implementation serves as a concrete example and ensures that data can be collected in a standardized fashion. The implementation must include both the quantum circuits defining the benchmark and the classical (post)-processing routines.

The evaluation of benchmark results demands rigorous statistical analysis. Therefore, alongside the benchmark definition, standardized evaluation code must be provided. This code should detail the statistical methods employed, including specific approaches such as weighting schemes for Gaussian, multinomial, or Bayesian statistics. To facilitate verification and future improvements, sample datasets should accompany the evaluation code. This enables automated testing and supports \textbf{continuous integration and deployment} (CI/CD) practices in the development of more efficient implementations. 

A critical aspect of benchmark definition is transparency. The implementation should explicitly document any error mitigation techniques employed~\cite{RevModPhys.95.045005}.
Even when the high-level implementation is open source (for instance through Qiskit), the limited visibility into hardware-specific implementations means that no one can truly guarantee the absence of error mitigation, compilation optimizations, or background enhancements at the hardware level. This fundamental opacity in the quantum computing stack presents a significant challenge for fair benchmark comparisons. It shall be noted that similar challenges are also present in classical computing.

Finally, quantum computing benchmarks inevitably involve classical computing components. This is particularly relevant in the NISQ regime, where the quantum part of any algorithm could, in principle, be fully simulated or fully enforced by classical computers. Therefore, benchmark definitions must clearly delineate and quantify the contribution of classical computing resources. This separation shall enable proper assessment of the quantum aspects of the implementation and provides a clear basis for evaluating potential quantum contribution.

The procedures defining benchmarks should, therefore, be
\begin{itemize}
 \item clearly defined,
 \item generic and SDK-independent,
 \item well-documented,
 \item transparent in resource use,
 \item separable from classical demands,
 \item scalable.
\end{itemize}
In the following, we summarize some of the benchmarking procedures suggested by researchers in the field.
\begin{itemize}
\item \textbf{Qubit-coherence benchmarking protocol}: It involves measuring the coherence time of qubits to assess how long they can maintain their quantum state before decohering.
\item \textbf{Entanglement benchmarking protocol}: Measures the ability of a quantum computer to create and maintain entangled states, which are crucial for many quantum algorithms.
\item \textbf{Noise characterization benchmarking protocol}: Involves systematic testing to understand the noise characteristics of quantum gates and measurements. This can include characterizing depolarizing noise, dephasing, and relaxation times.
\item \textbf{Randomized benchmarking protocol}: A protocol that evaluates the average fidelity of quantum gate sets by applying sequences of random gates and measuring the output. It helps robustly estimate the error rates without requiring full state or process tomography, even under state preparation and measurement errors.
\item \textbf{Circuit layer benchmarking protocol}: Evaluates the performance of quantum circuits based on the number of gates or layers, assessing how accuracy changes with increasing circuit complexity.
\item \textbf{Quantum error correction benchmarking protocol}: Assesses the effectiveness of quantum error correction codes in maintaining the integrity of quantum information over operations and time.
\item \textbf{Workload-specific benchmarking protocol}: Tailored benchmarking protocols that focus on specific quantum algorithms or applications, evaluating performance in realistic use cases.
\end{itemize}
Reviews on quantum computing benchmarking on various levels are available \cite{Proctor,eisert2020quantum}.
\subsection{Aggregating different metrics}

\begin{figure}
    \begin{center}
\centering\includegraphics[width=0.9\columnwidth]{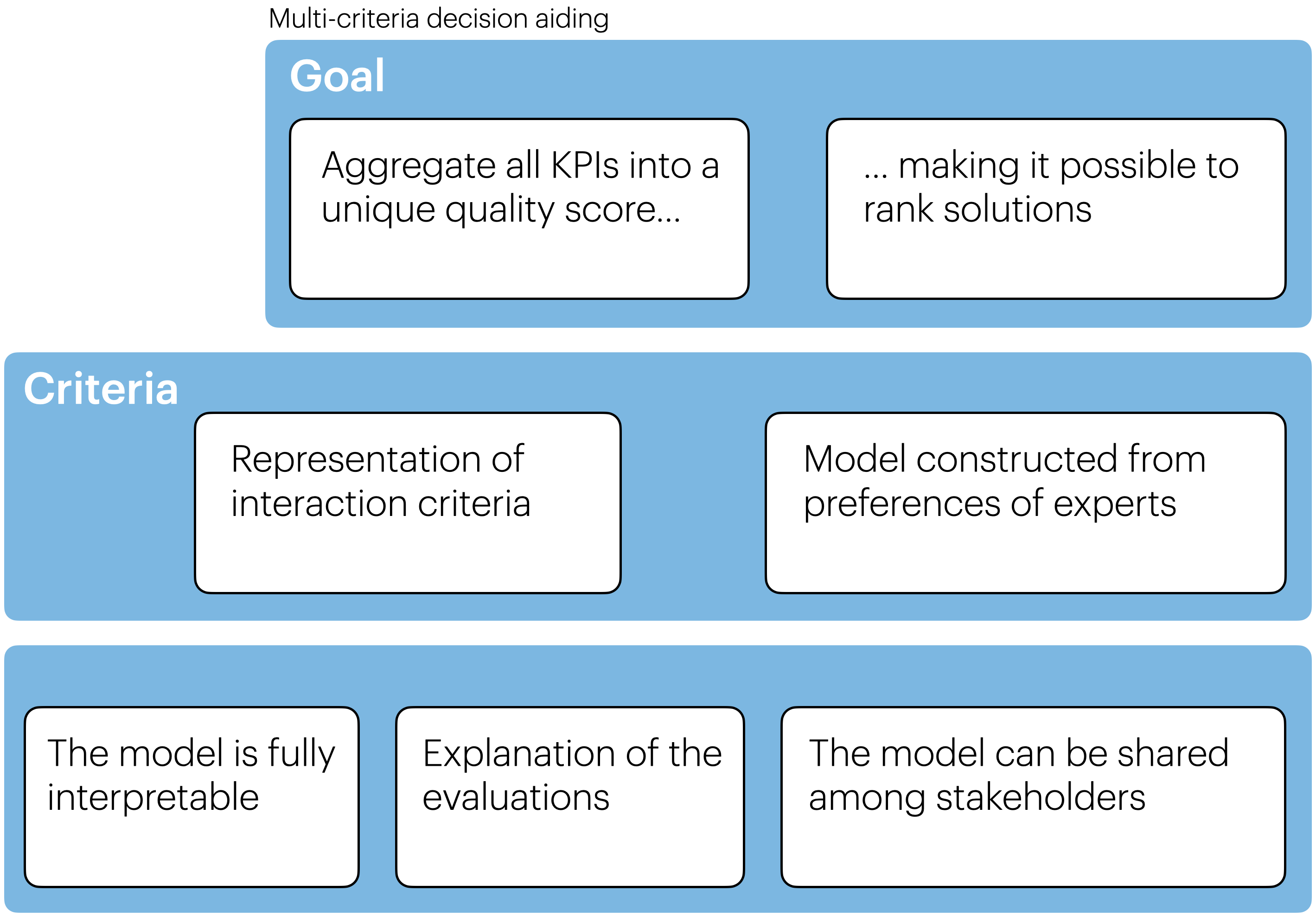}
    \end{center}
    \caption{Objectives for aggregations of metrics into an overarching metric.}
    \label{fig:aggregation_metrics}
\end{figure}

Additionally, recently, the need has arisen to combine multiple metrics in a weighted fashion to combine and rank multiple potentially conflicting metrics. For this reason, \textbf{multi-criteria decision aiding} (MCDA) methods have been suggested. MCDA methods are techniques designed to evaluate multiple, often conflicting criteria to aid decision-making and ranking. Each of these methods offers unique advantages and can be chosen based on the structure of criteria, the desired outcome, and the specific decision context. Objectives for aggregating metrics are provided in Fig.~\ref{fig:aggregation_metrics}.
Examples for MCDA methods are the following:
\begin{itemize}
\item \textbf{Analytic hierarchy process (AHP)}~\cite{Forman2001TheAH}: This hierarchical method involves structuring decision criteria and alternatives into a tree, assigning weights to each criterion through pairwise comparisons, and calculating a score for each alternative.
\item \textbf{Analytic network process (ANP)}~\cite{taherdoost2023analytic}: An extension of AHP, ANP captures dependencies and feedback among criteria, not limited to a hierarchical structure. It uses a network rather than a strict hierarchy to reflect interdependencies.
\item \textbf{Technique for order of preference by similarity to ideal solution (TOPSIS)}~\cite{Uzun2021}: Based on the concept of distance, TOPSIS ranks alternatives by their proximity to an ideal solution (closest to best, farthest from worst). Scores are assigned based on this relative distance.
\item \textbf{Multi-attribute utility theory (MAUT)}~\cite{keeney1993decisions}: MAUT assigns a utility value to each alternative by assessing how well it meets each criterion, factoring in risk and uncertainty. It maximizes expected utility based on the decision maker's preferences.
\item \textbf{MYRIAD}~\cite{labreuche2005miriad}: Multi-criteria methodology to evaluate and compare different solutions based on a set of conflicting criteria based on Choquet integral theory. Its main advantages are its ability to represent criteria that interact with each other, its learning modules of the model from simple information for a decision maker and its readability of the model. MYRIAD aims at capturing the preferences of experts.
\item \textbf{Elimination and choice translating reality (ELECTRE)}~\cite{Thakkar2021}: ELECTRE compares alternatives by outranking them based on criteria satisfaction. It does not focus on the absolute best but rather eliminates options that are less competitive.
\item \textbf{Simple additive weighting (SAW)}~\cite{ibrahim2019implementation}: SAW assigns weights to criteria, normalizes scores, and calculates a weighted sum for each alternative. It is straightforward but assumes independence among criteria.
\item \textbf{Weighted product method (WPM)}~\cite{Mateo2012}: WPM is similar to SAW but uses multiplication rather than addition. Each criterion score is raised to a power equivalent to its weight, which is then multiplied across all criteria.
\item \textbf{Data envelopment analysis (DEA)}~\cite{charnes1997data}: Process: DEA evaluates alternatives based on efficiency by comparing them with a “frontier” of best-performing options. It uses linear programming to measure relative efficiency.
\item \textbf{Preference ranking organization method for enrichment evaluation (PROMETHEE)}~\cite{Uzun2021}: PROMETHEE ranks alternatives based on pairwise comparisons of criteria, with scores generated to indicate the degree of preference of one option over another.
\item \textbf{Best-worst method (BWM)}~\cite{REZAEI201549}: BWM simplifies pairwise comparison by requiring only comparisons between the best and worst criteria relative to the others. This reduces decision-maker burden and improves consistency. 
\item \textbf{Compromise programming}~\cite{Ringuest1992}: This method identifies solutions that are closest to an ideal solution based on Lp-metric distances, aiming for a compromise when no single solution is entirely optimal.
\end{itemize}

\section{Different benchmarks}
\subsection{Component-level benchmarks}
\label{sec:component_level}

The smallest constituent of a quantum information device is a single qubit. Historically, the main method to characterise a qubit was based on the so-called $T_1$ time, a measure for how long populations remain unperturbed. In addition, the $T_2$ time describes how well superposition survive over time. These two numbers are typically provided across all quantum platforms. There are, however, at least three drawbacks: (1) both benchmarks have a unit, in this case, time, attached. Long time-scales may suggest superior performance, but these numbers would then need to be compared to, say, gate-operation times -- numbers, that are typically not accessible. In addition, (2) these two numbers are basis-dependent.
Finally, (3) the coherence times offer at most a lower bound on error rates, but barely allow one to make a predictive statement about the performance. 
For these reasons, \textbf{$T_1$ and $T_2$ times} are today 
less used (and will no further be 
considered in this document).

As is well-known, the dimension of the relevant vector space of a quantum register grows exponentially with the number of qubits. Thus, routines that try to describe states and processes in full detail will necessarily fail to be classically computable already at moderate sizes. \textbf{Quantum state tomography}~\cite{eisert2020quantum,ChuangNielsen97}  --  the recovery of unknown quantum states based on data  --  constitutes an important procedure for benchmarking for small systems, but is not scalable.
The same applies, to some extent, to readings of \textbf{gate-set tomography}~\cite{Nielsen+21}. For benchmarks to be scalable, they need to determine a small, potentially just a single number, e.g., the fidelity with respect to a state or process. Here, the much used \textbf{randomized benchmarking} suite with its many adaptations allows one to efficiently extract \textit{state-preparation and measurement} (SPAM) errors, as well as the error rate for gate operations~\cite{KnillBenchmarking, MagGamEmer, Mother}. Cycle benchmarking has been used to characterise multi-qubit gate operations reaching up to ten qubits. RB is commonly performed for gate operations that form groups, such as the prominent Clifford group, but also other finite groups. This is, however, rather a feature than a flaw, considering that Clifford gate operations are typically also the only operations that can be natively implemented on encoded qubits. What is more, a wealth of specific variants of randomized benchmarking schemes are known \cite{Mother}. Therefore, RB can be readily applied onto logical qubits and offers the required compatibility from the NISQ to the FTQC regime (in which one can also consider RB on the logical level). Based on RB, one can estimate key benchmarks such as the average gate fidelity in a SPAM-robust fashion, rendering it one of the most important procedures for benchmarking quantum gate sets. 

While the full state cannot be efficiently recovered both concerning sample and computational complexity, many properties, in fact, can. The term \textbf{property testing}~\cite{Flammia-PRL-2011,Hangleiter,KlieschRoth,Leandro,PropertyTestingReview} refers to the notions of verifying in a reliable and robust fashion whether a 
desired property is present in an unknown quantum state used in and for a quantum computation. This could be the
fidelity with a desired target state, a level of entanglement  or a degree of so-called magic as a computational resource. These studies are rooted in
\textbf{learning theory} and give rise to schemes that are usually sample and computationally efficient, while quantum state tomography is not.

A more challenging question is to describe not only individual components, such as single gate operations, but potential \textbf{correlations} in the system: Is there cross-talk from a gate acting on one qubit onto another qubit? Do measurements on one qubit affect other qubits in the system? And if so, how? The effects may be non-local or non-next-neighbour and thus challenging to characterise in a scalable fashion. One of the few methods to detect and characterise correlations is \textit{averaged circuit eigenvalue sampling} (ACES) \cite{ACES,Hockings:2024xyz}.
Also, one can perform cross-talk tomography based on RB data \cite{RandomSequences}. 

An increasingly important development is that of benchmarking protocols of quantum error correction protocols. Indeed, one can perform RB on the logical level, and can hence calibrate and benchmark the performance of quantum gates on the logical level of a quantum error corrected quantum computation.

FTQC systems require multiple physical qubits to represent a single logical qubit. The \textbf{QEC overhead} quantifies the number of physical qubits required per logical qubit, which is a key metric in evaluating the scalability and efficiency of FTQC systems. A low QEC overhead with a high logical density, which is the ratio of logical qubits to physical qubits, is desirable for ensuring that the FTQC system can scale effectively without becoming prohibitively resource-intensive.

Additionally, FTQC systems require the ability to perform quantum operations in real-time, which introduces additional metrics for evaluating the system’s performance during fault-tolerant computations. The ability to perform mid-circuit measurements and reset qubits during an ongoing computation is crucial for FTQC. This metric evaluates how efficiently a quantum system can interact with qubits in the middle of quantum circuits, without introducing errors or disrupting the state of the system. It is particularly important for real-time quantum error correction and dynamic adjustments during the execution of quantum algorithms. It shall be mentioned that there is recent progress on investigating measurement-free FTQC, which may reduce the technically challenging demands on mid-circuit measurements and reset~\cite{https://doi.org/10.48550/arxiv.2410.13568}.

As FTQC relies on continuous error correction, the decoding throughput becomes a vital metric. It measures the system’s ability to detect and correct errors at high speed without affecting the execution of quantum operations. \textbf{High decoding throughput} ensures that error correction cycles can be completed efficiently even in larger systems, allowing FTQC systems to maintain coherence and reliability during computation.

The \textbf{logical clock rate} measures how fast logical qubits can be updated, while the full-decoding response time assesses how quickly the error correction process can be completed in its entirety. These metrics are essential for understanding how well FTQC systems can operate in real-time, ensuring that logical qubits remain accurate and error-free even as the system scales and the error correction process becomes more complex.

\subsection{System-level benchmark}
\label{sec:system_level}

One of the more prominent holistic benchmarks for quantum information processors is the so-called \textbf{quantum volume}~\cite{Cross+19}. It provides a statement whether a certain number of qubits and gate operations provide a certain average success probability. The calculation of the classical output probability distribution becomes exponentially difficult with number of qubits and gate-depth, and therefore does not fulfil our benchmark requirement to be scalable. What has been proposed to replace the quantum volume, is the \textbf{error per layered gate}
\cite{AtScale,LayerFidelity}, measuring  how errors propagate through an architecture, in a more scalable fashion. It provides a benchmark that encapsulates the entire processor’s ability to run circuits while revealing information about individual qubits, gates, and crosstalk.

Within Europe, a team is currently adopting the ideas of the quantum volume to realise the \textbf{Clifford volume} 
Here, the gates are restricted to Clifford gate operations. Thus, the output state can be calculated, and compared, in a classically efficient fashion. The benchmark provides therefore a reasonable predictive power about the quality of gate operations for arbitrary system size and depth. In addition, the benchmark can be readily implemented both in the NISQ as well as FTQC domain.

One of the early benchmarks to evaluate multi-qubit control in quantum information processors has been  --  and still is, e.g., when recently
a 120 qubit GHZ state has been prepared on an IBM
Heron R2 quantum processor 
\cite{Heron}  --  the \textbf{creation and evaluation of GHZ states}~\cite{TothGuhne05}. Both the populations as well as the coherence of the GHZ state can be readily evaluated in a scalable fashion, using a sequence that involves local and entangling gates across the entire quantum register. Another advantage of the GHZ state evaluation approach is its compatibility with the requirement to be applicable both in the NISQ domain, acting on physical qubits, as well as in the FTQC domain, acting on logical qubits. It shall be noted that the requirements for extending GHZ state benchmark into the FTQC domain requires ‘single-shot’ readout to be implemented and directly evaluated. For that reason, methods such as readout-error-correction shall not be employed in the context of this measure, as it does not scale into the FTQC domain.
It shall be noted that Clifford volume and GHZ state fidelity are both implemented on the instruction level of the device and thus bypass aspects such as the quality of the employed compilers.

There are numerous methods to improve system-level performance by different forms of error mitigation. These methods include optimisations on the compiler level (e.g., hardware specific optimisation to reduce the overall gate count). In addition, these compilers may take into account more detailed gate characterisations and include effects such as crosstalk to optimise the overall performance including such effects. In addition, one can implement error mitigation methods such as dynamical decoupling to reduce static or slowly drifting parameters in the setup. Alternatively, one can implement randomized compiling to suppress systematic errors. While the compilation of circuits is numerically challenging, the integration of dynamical decoupling as well as randomized compiling is relatively straightforward and can be used both in the NISQ as well as in the FTQC domain. As such, the \textbf{usage of these mitigation methods} are entirely at the end-users discretion in the context of the system-level benchmarks and higher levels of abstractions, such as HPC-level benchmarks or application-benchmarks. 

If system-level benchmarks are out of reach, \textbf{cross device verification} can in instances  still be available \cite{PhysRevLett.124.010504,PhysRevA.107.062424}. This refers to a body of work that allows to verify the performance of 
a quantum system if an efficient classical simulation is out of reach. Oftentimes, this cross device verification will be sample and computationally efficient \cite{Cross}.

There are upcoming benchmarks that also try to cover the speed at which quantum gate operations can be executed on a device. These benchmarks would cover aspects such as the possibility to implement gate operations in parallel or may shed light on details about the time it takes to implement a gate vs the time it takes to first rearrange the quantum register to support the execution of the targeted gate operation. Here, topology and connectivity play a significant role. One benchmark to cover these aspects would be the \textit{reliable quantum operations per second} (rQOPS), suggested by the Microsoft team. Unfortunately, there does not seem to be a publication available that clearly describes the intended procedure. Alternatively one could use the already above mentioned \textbf{circuit layer operations per second} (CLOPS)~\cite{LayerFidelity,Bench-QC}, suggested by IBM. Here, the CLOPS benchmark can be extended from NISQ towards FTQC, which will subsequently provide meaningful estimates on time-to-solution with beneficial QEC.

The evaluation of speed in the context of QEC will become more important in the context of QEC, considering that syndrome extraction and classical processing to determine the error is in principle NP hard~\cite{PhysRevA.83.052331} in worst-case complexity. Here, CLOPS in the FTQC regime are expected to highlight the requirement for low-latency interfaces between quantum computers and classical computational
solutions  --  potentially directly with HPC systems.

Incorporating \textit{fault-tolerant quantum computing} into system-level benchmarks requires a nuanced understanding of error correction protocols and their performance metrics. Beyond traditional benchmarks like 
the quantum volume and Clifford volume, which provide insight into system performance in the \textit{noisy intermediate-scale quantum} domain, FTQC requires more specialized metrics that account for the resilience and reliability needed to achieve fault tolerance.

One critical metric for FTQC is \textbf{quantum operations} (QuOps) which tracks the number of quantum operations that a system can reliably execute, including gate operations and measurements. The reliability of these operations is essential for ensuring that quantum computations can proceed without error accumulation that could invalidate the results. These benchmarks are still evolving, and detailed publications describing their procedures remain scarce.
Moreover, CLOPS can be extended into the FTQC domain, providing meaningful insights into time-to-solution when beneficial \textit{quantum error correction} (QEC) methods are employed. 

In FTQC, one of the most pressing concerns is the \textbf{efficiency of QEC}. 
As logical qubits are constructed using multiple physical qubits to protect against errors, FTQC systems must efficiently manage this overhead. The \textbf{Lambda parameter~$\Lambda$} \cite{lambda1,lambda2} plays a critical role in assessing QEC efficiency, as it measures the rate at which logical errors are suppressed in systems with increasing numbers of qubits. A low Lambda suggests a robust error correction process capable of scaling without introducing significant overhead. As qubits and gate fidelities improve, it becomes increasingly important to assess whether the error correction methods employed are both effective and resource-efficient.

Finally, the concept of \textbf{MegaQuOp} -- a benchmark proposed~\cite{megaquop} for large-scale quantum systems -- measures the collective performance of a quantum computer as it handles millions of quantum operations in a fault-tolerant regime. As FTQC systems scale, reaching MegaQuOp thresholds will represent a significant achievement, marking the point at which quantum computers begin to surpass the performance of traditional supercomputers in solving complex, previously intractable problems. Achieving such benchmarks will indicate that FTQC systems are not only theoretically viable but also capable of providing practical value, especially as industries seek to leverage quantum computing for real-world applications. This can complement quantities
such as the \textit{average detector likelihood} to benchmark
QEC schemes \cite{QECBenchmarking}.

These FTQC-specific metrics -- QuOps, rQOps, CLOPS, Lambda, and MegaQuOp -- offer a more comprehensive suite of benchmarks, providing a clearer picture of how quantum systems evolve from their noisy intermediate-scale implementations toward robust, error-corrected fault-tolerant devices. Such advancements are essential to realize the potential of quantum computing in solving complex problems across fields like materials science, cryptography, and optimization.

\subsection{Software-level benchmarks}
\label{sec:software_level}

Software permeates the entire control stack through several software layers and APIs to communicate the desired operations from the end-user to the actual execution on a quantum processor, from the instruction set sent to the quantum processor, via computer-internal automation and calibration, to data processing and data provision. Benchmarking of software aspects within a quantum computer is therefore non-trivial. In the context of this document we shall focus on aspects that \textbf{improve the performance of a quantum computing} device.

At the lowest levels, compilers try to map a target unitary in an efficient fashion onto a hardware, given its capabilities in terms of coupling topology, quantum gate operations, readout processes, and semi-classical control. Note that the exact compilation of quantum circuits is NP-hard~\cite{Botea2018OnTC} in worst case complexity. Therefore, the compilation is expected to become more and more resource-intensive 
as the system size grows in the number of qubits and circuit depth, rendering an optimal compilation difficult.
Here, a \textbf{quantum compilation volumetric benchmark}, similar to quantum volume, that states how many qubits and how many layers of gate operations can be compiled (e.g., in a given time) may offer insight into the scalability of quantum devices in the near future. To date there are no volumetric benchmarks that cover the size that hardware-optimised compilation routines can cover. These volumetric compilation benchmarks would require standardised target-computations to be realised. Here we could build on top of standardised quantum applications, such as Shor’s algorithm, to evaluate the performance as the size of the problem and system increases.

If software-level benchmarking is out of reach, \textbf{cross-device benchmarking} may still be possible. Even if the efficient simulation of a quantum device is infeasible computationally, one can still benchmark one quantum device against another. This approach is pursued in cross-device benchmarking~\cite{Cross}. 

Considering that large-scale quantum computation at sufficiently low error rates will require quantum error correction, it is also important to note that \textbf{error syndrome mapping} onto the error source is an NP-hard problem~\cite{PhysRevA.83.052331} in worst case complexity. Considering that the error correction should happen in real-time (meaning, on time-scales similar to quantum gate operations), there are several important figures: (1) what size of syndrome can be evaluated on the time scale of a gate operation (between~100\,$ns$ for superconducting systems to~100\,$\mu s$ for trapped ions, depending on the platform)? If heuristics are used to meet those time-scales, how much will approximate solutions negatively affect the overall performance of the quantum computer? There are currently no established benchmarks to cover the speed and quality of error syndrome mapping.

\subsection{HPC-level and cloud benchmarks}
\label{sec:hpc_level}

The design of robust and comprehensive benchmarks for computer hardware is an ongoing challenge, and it has been a focal point of research in the \textit{high-performance computing} (HPC) community for decades. The technical issues encountered during the development of early HPC benchmarks closely parallel those now faced in the realm of quantum computing performance evaluation. Historically, classical computing performance was often quantified through HPC system benchmarks, which provided comparative analysis across different machines using low-level performance indicators, such as \textbf{instructions per second} (IPS). While these metrics, along with others like memory bandwidth and \textbf{floating-point operations per second} (FLOPS), offered granular insights into various components of a machine’s architecture, they required the development of higher-level, more interpretable metrics to facilitate broader comparisons and practical utility assessments. Lessons learned from the HPC domain provide valuable guidelines for establishing effective benchmarks that can achieve widespread adoption and accurately represent system capabilities.

One of the most enduring and widely accepted benchmarks in classical computing is the \textbf{LINPACK} benchmark~\cite{dongarra1979linpack},~\cite{Linpack_2} and its successor LAPACK~\cite{LAPACK}. It assesses system performance by solving dense systems of linear equations, a task with high computational demand. Despite its prominence, there has been significant discourse within the scientific computing community regarding whether LINPACK’s focus on linear algebra accurately reflects the diverse range of real-world workloads encountered in scientific and industrial applications. Notably, the creators of LINPACK did not intend for it to serve as a universal performance metric, and they have acknowledged the inherent complexities in benchmarking. These complexities arise from a multitude of factors, including algorithm selection, software implementation details, hardware architecture, and the extent of optimization efforts applied by human developers. Each of these elements can have significant impacts on the measured performance, leading to potential disparities between benchmark results and real-world application performance.

The reliance on simplified metrics like LINPACK persists, in part, due to the desire for straightforward comparison between HPC systems. The TOP500 list, which ranks supercomputers solely based on their LINPACK performance, exemplifies this trend~\cite{Top500}. However, the limitations of LINPACK in representing more diverse computational tasks have spurred the development of complementary benchmarks, such as the \textbf{HPCG)} (\textit{high performance conjugate gradients}) benchmark~\cite{HPCG}. HPCG is designed to better align with performance characteristics seen in typical scientific applications, such as sparse matrix computations, which are more reflective of workloads in fields like physics and engineering. Even with well-established hardware architectures, developing a single, all-encompassing performance benchmark remains a highly complex task due to the inherent diversity of computational tasks and system architectures.

As ``high-level'' benchmark, based on a use-case, such as LINPACK (which focuses on linear algebra, as stated above), another lesson learned from HPC is the need for getting information about the performance of each part involved in the computation. A benchmark's result can be easily biased by some side effects, but this can be worked around by adding other, more hardware or application specific, benchmarks. However, measurement of individual components of the system, rather than the overall performance of the entire system may lead to completely different conclusions. An example is performing a benchmark in order to get the available performances for a distributed and parallel filesystem. The available network bandwidths, the network topology, the memory performance of the servers, or the IO rates of the involved disks are to be measured independently, by dedicated tools. 

In summary, while classical HPC benchmarks like LINPACK provide a useful, albeit narrow, view of system performance, the ongoing refinement of benchmarks, such as HPCG, highlights the necessity for metrics that more comprehensively capture the multifaceted nature of hardware performance across a range of real-world applications. This approach serves as a valuable reference in the ongoing efforts to establish meaningful and effective benchmarks for emerging technologies such as quantum computing.

Besides lessons learned from HPC benchmarking, there is another dimension to quantum computers benchmarking that relates to HPC. Quantum computers are increasingly being integrated into HPC systems, and a growing number also becomes available through cloud platforms. In cloud environments, resource allocation and user management may significantly differ from traditional HPC access protocols, making it essential to develop suitable benchmarks for evaluating performance. Unlike HPC-based quantum setups, cloud benchmarks must account for the unique nature of cloud access to the QC, focusing on metrics such as computational power, storage speed, memory throughput, network latency, and instance provisioning time. Additionally, to reflect real-world conditions, benchmarks should consider queue times and cloud front-end overheads, especially for hybrid HPC-QC configurations. Metrics like \textbf{mean instance provisioning time} and \textbf{IOPS} (input/output operations per second) are valuable for assessing performance in data-intensive, hybrid classical-quantum applications, where efficient cloud-based QC is crucial.

Certain metrics established for hybrid HPC-QC systems, such as \textbf{average queue time}, are also applicable to cloud-based QC. However, queue time is influenced by multiple factors, including the resource management settings of the HPC system, system availability, and the size of computations submitted by other users. As a result, while this metric is important for end-users, it reflects a blend of factors shaped by the vendor’s infrastructure, the service provider’s resource management, and the operational patterns of other users.

Some cloud providers offer access to exclusive time windows for computations. Although these dedicated blocks promise uninterrupted computational time, latency can still occur due to factors such as queuing delays, automated system calibration, and time consumed by classical data processing.
Given these considerations, various latency-related benchmarks developed for hybrid HPC-QC systems may be applicable to cloud-based quantum devices, allowing us to re-purpose established HPC-QC benchmarks for evaluating cloud-accessed quantum resources.

\subsection{Application-level benchmarks}
\label{sec:application_level}

Application-level benchmarks consider the performance of a complete quantum computing software stack together with quantum hardware for specific application use cases. The use cases need to be clearly defined in a mathematical format, usually together with the input data. Typically, different (quantum) algorithms, together with different (quantum) software and hardware can be used to provide a solution to the use case specified. Consequently, different metrics are employed to measure the performance of the full solution pipeline and are combined into an application-level benchmark.

Due to the wide applicability of quantum devices, and due to the fact that application-level benchmarks require specific problems to be considered, a wide variety of application-level benchmarks (and specific metrics to measure the performance of certain aspects in the computation) can be defined, and have been defined. Moreover, many algorithms have known outcomes, which provides a scalable method to assess the fidelity of the quantum computer. Generally speaking, an application-level benchmark is completely defined by the problem the benchmark considers, the KPIs and metrics that need to be measured during solving of the problem, and a protocol. 

Interestingly, there is only a small number of quantum algorithms  --  less than a 100 --
known to outperform their classical counterpart \cite{MontanaroOverview}, where only a small fraction of these algorithms are used in most quantum programs. Considering the notably slower clock-speed of quantum computers compared to classical computers (between three to six orders of magnitude), we will focus on those applications based on algorithms offering at least a superpolynomial or ideally exponential speed-up. This reduces the algorithms to about 30 algorithms with superpolynomial speedup, and three with exponential speedup. Note that the superpolynomial speedup of these algorithms is largely based on heuristic considerations and
is actively investigated and disputed in the community. This significantly narrows, but also focuses, the efforts of quantum applications, where even minute improvements on the system side may result in significant gains compared to classical capabilities.

Historically, a significant motivation for building quantum computers had its origin in the development of a \textbf{quantum-efficient factoring algorithm} by Peter Shor in 1994~\cite{shor1994}, which offers a superpolynomial speedup compared to known classical methods. Structurally, the algorithm covers several aspects suitable for establishing a quantum benchmark: (a) the use case is clearly defined. (b) Its structure includes the quantum Fourier transform, which also plays a role in other algorithms such as quantum phase estimation as a subroutine in algorithms for quantum chemistry. Finally, (c) the algorithm can be implemented in the NISQ domain, but will have to be implemented in the FTQC domain to offer sufficient computational accuracy to factor large numbers. For these reasons, a scalable implementation of the Shor algorithm is also one of the European Quantum technology flagship KPIs \cite{EUFlagshipKPI}.

Quantum computing has been proposed to provide solutions to industrial-scale optimization challenges, although those are typically NP-hard problems. Consequently, a lot of attention has been put on the so-called \textit{quantum approximate optimization algorithm} (QAOA) \cite{QAOA,Lukin}, which suggests applications in optimization routines such as Max-Cut, even though no proven advantage over classical algorithms has been shown yet. 

An example for an application-level benchmark addressing optimization problems is the \textbf{Q-score}~\cite{QScore}, as originally proposed by Atos, which measures the performance of gate-based quantum devices at solving the Max-Cut problem. It measures the KPI's problem size and obtained solution quality. It then defines the Q-score as the largest problem size for which the obtained solution quality is significantly better than a random solution over a large enough number of random samples. This is determined by using scalable approximations for both the optimal solution and random solution for the given problem size. The Q-score has seen quite some follow-up work, with authors extending the Q-score to quantum annealers, Gaussian Boson samplers and classical devices, authors imposing a time limit in which the solution needs to be obtained, and authors defining the Q-score for vastly different problems. The most recent work of van der Schoot et al.~\cite{van2024extending} details how the Q-score can be seen as a framework that can be utilized to benchmark quantum devices with any problem (given a few lenient restrictions), with different degrees of freedom (such as runtime and amount of optimisation) left as a choice for the end user.

Similarly, the \textbf{QPack}~\cite{QPack} framework yields a set of quantum metrics designed to benchmark gate-based quantum devices. It can be instantiated with various problems, examples being the Max-Cut Problem, the Dominating Set Problem and the Travelling Salesman Problem. For obtaining a benchmark, the problem size, obtained solution value, problem runtime all need to be measured. From this, the accuracy of the solution, and a quantity called the scalability are computed. This then yields four different KPIs, namely the problem size, accuracy, runtime and scalability. These four are rescaled and plotted in a radar diagram. Each of these four rescaled values, as well as the area of the shape in the radar plot yield the metric output for the used quantum device.

The \textbf{quantum application score} (QuAS)~\cite{Mesman+24} combines the efforts of the Q-score and QPack benchmarks. The result is a application-level framework in which a user has many degrees of freedom to benchmark a quantum device. In this way, a user can optimally choose what is relevant for the specific application and hence choose the most suitable quantum device. To use the QuAS, a user first has to choose a specific problem to benchmark quantum devices with. In addition, the user chooses which KPIs are important, and in addition chooses the importance (weight) of each of these KPIs. The quantum device is then used to solve various instances of this problem and for each the required KPIs are measured. The QuAS then combines these measured KPIs in an intricate way to return a single score detailing the performance of the quantum device on the given problem with the specific KPIs in mind.

However, despite the numerous efforts in defining application-level benchmarks addressing optimization problems, these efforts need to face the challenge that there has been progress in the field of classical algorithms, to the point it is not clear whether the benefit from QAOA compared to classical solutions may only be polynomial. Subsequently, several benchmarks based on QAOA and Max-Cut (e.g. Q-score) may not offer any insight into the capabilities of quantum computers. We will therefore \textbf{refrain from suggesting QAOA and its derivatives for quantum application benchmarking}.

The QAOA belongs to a larger group of algorithms, the so-called \textit{variational quantum algorithms} (VQA)~\cite{Variational}.
These have been recently seen in the field of programmable sensors, chemistry, and more. VQA get praised for its divide-and-conquer approach: a problem gets separated into a (1) quantum computation based on a classical parametrization, while (2) a classical computer optimizes the parameters such that (1) finds a global optimal for the problem at hand. While this structure has found numerous proof-of-concept implementations, the community has later learned that the classical parameter optimisation is, generally, NP-hard \cite{Bittel} in worst case complexity. 

Going beyond the specific class of optimization problems and variational algorithms, the community proposed various wider benchmarks. One of the first application-level benchmarks comes from the \textbf{Quantum LINPACK}~\cite{QLinpack}. The authors take the state of the art of high-performance computing benchmarking, named LINPACK, and suggest a quantum alternative suitable to benchmark quantum devices. The benchmark measures how well a device performs at implementing a certain model called the RAndom Circuit Block-Encoded Matrix (RACBEM) model. This model can be used to perform a wide variety of linear algebra tasks. The underlying metric can hence be seen as a measure for linear algebra performance. Performance of the metric is measured by the probability of measuring a certain state, which measures the accuracy of the implementation of the model. This probability directly yields the metric score.

Various initiatives propose benchmarking suites consisting  of multiple application-level benchmarks and addressing a series of (industrial) use cases. An example for this is the \textbf{QED-C metrics}~\cite{lubinski+23,lubinski+24}. While this originally started as a single application-level metric framework, it has now been built out to a large benchmarking endeavour covering various approaches and metrics. Within this endeavour, various problems are used to benchmark quantum devices, ranging from tutorial algorithms such as Deutsch-Jozsa and the Quantum Fourier Transform to real-world problems like Shor’s algorithm and the Max-Cut problem. For these benchmarks, the KPIs problem size, solution fidelity and runtime (defined in various ways) are measured. The earlier benchmarks had as output the solution fidelity (as compared to a perfect execution of the quantum circuit) for different problem sizes. Later benchmarks focus more on the solution fidelity (as compared to the best possible solution) and runtime for different problem sizes. The output of these benchmarks is always given in a visual way, with colours showcasing the (quality of the) fidelity/runtime for the different problem sizes.

Similarly, the \textbf{SupermarQ}~\cite{Tomesh+22} framework aims to yield a scalable, hardware-agnostic, application-level quantum benchmarking framework, inspired by techniques from the classical benchmarking domain. It evaluates the performance of quantum devices on eight different mathematical problems and quantum algorithms, while not extending to application use cases. The quantum algorithms and quantum states to prepare range from GHZ-state preparation and Mermin-Bell to different flavours of QAOA, ignoring here if these algorithms may reach an superpolynominal speed-up or not. For each of the quantum algorithms, a different score is computed for different problem sizes (or equivalently, for different number of qubits). These scores yield the benchmark output. Each of the scores is a value within~0 and~1, and each of the scores roughly correspond with the accuracy with which the quantum device implements the considered algorithm. 

\textbf{BACQ}~\cite{Barbaresco:2024fmg}, as currently developed in a dedicated project, is a package for multicriteria benchmarking of quantum computers performance at the application level, intended to make sense for industry end users. The package encompasses 
\begin{enumerate}
\item reference problems to be solved, related to simulation of quantum physics (one-dimensional 
Ising model with transverse field, two-dimensional Hubbard model, BCS model etc), 
combinatorial optimization (Maximum Cardinality Matching / Gn series, MaxCut, quantum walks etc), linear system solving, and prime factorization,
\item technical metrics to evaluate the resolution of these problems, as various as computation time, problem size, accuracy, fidelity, energetic consumption, Q-score, and more,
\item a model for aggregation and analysis of these metrics, providing an overall notation of the quantum computer, taking into account high-level operational indicators.
\end{enumerate}
In particular, regarding Q-score, the metric was originally defined to evaluate the resolution of the MaxCut problem~\cite{QScore}, as the largest size of the graph on which the problem can be solved with good approximation. This first version is open-source, available on several GitHubs, and has been implemented on a large variety of quantum computers. In the project, the methodology is extended to the resolution of many-body problems among others. About energetic consumption assessment, the framework developed is based on the description of resource efficiency as the ratio of performance metrics to resource cost, considering a full stack approach of quantum computers and a methodology called MNR (\textit{metric noise resource})~\cite{fellous2023optimizing}. Regarding metrics aggregation and analysis, the MCDA tool is developed and includes metrics identification, normalization and aggregation using Choquet integral, as well as notation elaboration and explanation. It takes into account interacting metrics and user preferences and provides a fully interpretable and explainable notation. The final notation is to allow comparisons between different quantum computers and with classical computers. The whole package is designed to cover analog to gate-based quantum computers, from the NISQ to FTQC regimes. It aims to be as open as possible and, particularly, able to take into account the evolution of quantum computing technologies as well as of performance evaluation benchmarks and metrics. This is key to make it widely recognized and used by quantum computing stakeholders. The main goal of BACQ is to provide objective, transparent, reproducible and comparable performance evaluations, allowing reliable assessment of progress towards useful quantum computing.

The German \textbf{Bench-QC} project~\cite{Bench-QC} develops a systematic benchmarking pipeline for the areas of industrial simulation, optimization and machine learning use cases with the intend to enable a well-founded potential analysis for the future industrial use of quantum computing. To achieve this mission, the project defines various use cases coming from these three fields. Examples include simulated interaction and dynamics of fermionic ions, sensor placement in water networks, generative design for construction elements and crack detection. These industrial use cases are mapped into their mathematical formulation. Sequences of different problem sizes and increasingly difficult problem instances are formed. For the resulting instances, a classical baseline is implemented and compared to different suited quantum algorithms. The performance of the selected quantum algorithms is both simulated and measured on different quantum hardware, here comparing in particular the performance of quantum computers based on superconducting qubits as well as on ion traps. To evaluate the performance of both classical and quantum algorithms, different metrics are used, which span from characterizing the quantum hardware used with metrics such as the quantum volume, the coherence time, the gate fidelity etc. to estimating the resources used by the algorithms such as e.g., the learning steps required. The overall performance is measured for example by the overall accuracy reached by the algorithm. Technically, the benchmarking procedure is realized by the modular \textbf{QUARK} framework~\cite{Finzgar:2022mya} with standardized and reproducible pipelines. This framework in particular also specifies the data used, as well as any classical transformation step or optimizer required on top of the quantum algorithms and the software used (such as e.g., Qiskit
or Pennylane.

The EuroQCS-Poland project has developed the \textbf{Open QBench} benchmark suite, which includes several quantum algorithms commonly used on NISQ devices, particularly to evaluate the performance of new quantum computers to be established and supported by EuroHPC JU in 2025~\cite{QBench_soft}. It leverages established methodologies from high-performance computing benchmarks to identify characteristics that make a benchmark both useful and widely applicable for end-users and quantum application developers~\cite{QBench}. The QBench benchmarking framework has already been tested on various quantum devices, including Josephson junction-based superconducting qubits, trapped ions, quantum annealers, and Boson sampling, to measure their performance. Recently, QBench adopted a Multi-Criteria Decision Analysis method to rank NISQ devices based on their comparative performance. Finally, the proposed QBench benchmarking approach can be easily extended with new quantum algorithms and applied to evaluate the performance of emerging and upgraded quantum computing platforms.

The mentioned QUARK framework is one of the very few examples offering a systematic benchmarking pipelines for \textit{quantum machine learning} (QML)
\cite{biamonte2017quantum}. QML is an intersection of machine learning with quantum computing, where parametrized quantum circuits are trained and adapted to learn from either classical or quantum data. Similarly to classical machine learning, QML algorithms are typically empirically benchmarked by looking at the final reached accuracy or loss (or other metrics known from classical machine learning).

\subsection{Additional classical metrics to consider}
\label{sec:classical_metrics}

Besides the quantum-focused characterization methods described above, the classical aspects of a quantum computer shall not be ignored either, as they offer insight into the challenges to overcome as the systems are scaled up. Here, aspects such as the volume of the system, its power consumption, but also the dependence on the supply chain and its sustainability play a continuously more important role as the systems mature.

The volume of the system plays an important role to describe the number of components as well as the level of integration. Note that the volume should include not just the processor, but all enabling technologies (e.g. RF and MW sources for cryo-systems), or detectors (e.g. in the case of photonics systems). Note that for several technologies the number of qubits may significantly increase, without the volume of the device growing. However, additional control may and likely will result in increasing system size. As such, \textbf{volume per qubit} may be an interesting number to track for the years to come.

In addition to the volume, the total power consumption of a system, and subsequently watts consumed per qubit, will become an interesting number. Here, the EU sustainability act and similar measures will start to play an important role in terms of \textbf{power-to-solution}, complementing \textbf{time-to-solution} that should be tracked for the execution of most of the benchmarks stated within this document.

The document highlights that benchmarks are ideally applicable both in the NISQ as well as in the FTQC domain. Subsequently, statements based on energy consumption per qubit, volume per qubit, are ambiguous as logical qubits may consist of several physical qubits, and logical computation consumes magic states (based on physical qubits) during the execution. One may think that a transition towards power consumption per computation may be easier to define. However, with the rise of hybrid systems, this results in further open questions: shall the computational time on a classical device be included? How do we compute the power consumption of hybrid systems, in particular if one of them may be idling? Also note that the power consumption may be constant or highly dynamic, and differ from classical to quantum devices. With growing maturity of these hybrid systems, such questions can be addressed.

Another aspect to consider is the usage of \textbf{sustainable materials}. For instance, several solid state solutions require cryogenic temperatures, partially below 4\,K. For this temperature range typically dilution refrigerators are used that require ${}^3\text{He}$. However, ${}^3\text{He}$ is a non-replenishing commodity, that is produced mainly for military purposes or national security at facilities outside the EU. This raises the question about the technology \textbf{supply chain} within quantum technologies, and whether these systems can be developed, produced, and maintained from local resources. The example above highlights that, at least some technologies, currently have to rely on non-European suppliers.

Finally, the \textbf{system life-cycle} of the system ought to be considered beyond the realisation and operation. Some of the envisioned systems may use radio-active isotopes that may offer benefits for the quantum technology, may also affect the post-use recycling and disposal of quantum technologies.

\section{Standards}
\label{sec:standards}

Standardisation supports progress in the quantum community by offering a unified language and definitions. National and transnational organisations, such as ISO or CEN-CENELEC are currently working on documents to ensure that users can rest assured that ``quantum inside'' really corresponds to the usage of a quantum device, and may not mean that one is using classical computer that is emulating a quantum device.

Focusing on quantum benchmarking, standardisation is meant to establish protocols to provide reproducible, transparent, and comparable results. Here, the standardisation activities can build on established routines from classical software engineering and standardisation in that field: (1) an open protocol with unambiguous instructions on how to implement the necessary operations, (2) accompanied with standardised evaluation routines to ensure that data is not only generated but also evaluated in a standardised fashion, and ideally (3) would host the information at a location that would allow comparison and cross-checks of statements at a later date.

Regarding quantum technologies, as any other emerging technology, standardization supports technology development, industry uptake, market development and social adoption by providing reliability, consistency, compatibility and commonality. More precisely, standardization, which is based on consensus, supports establishment of common language, common practices, interoperability, to name a few. This is key to build a consistent ecosystem, including researcher, investors, industry, market and society; to build a consistent and secured supply chain; to ensure integration of quantum technologies with the existing ones; to enable market access to vendors; to give trust to quantum technology users. In addition standardization is also often considered as a tool for influence and technological power. Practically, standardization of quantum technologies is challenging, because of the often low level of maturity of these technologies. The standardization objects shall be adapted to the TRL of the quantum technologies of interest, in order to avoid standards that would hinder innovation. Thus, depending on the TRL, standardisation addresses vocabulary, terminology, key characteristics definition and measurement, benchmarks, performance assessment, interfaces, interoperability, products, risks and security, and more. The standardization of the products of services per se should be considered once performance have been demonstrated. A recent study led by EU has demonstrated that the current need for standardization mainly concern benchmarks for performance evaluation. 

Indeed, benchmarks for quantum computers are included in the European CEN-CENELEC standardization roadmap on quantum technologies, which first release was published in March 2023, after preparation by the pre-standardization CEN-CENELEC Focus Group on Quantum Technologies (FGQT). Consistently, within the CEN-CENELEC JTC22 on ``Quantum technologies'' established in 2023, Working Group 3 on ``Quantum computing and simulation'', a preliminary work dedicated to QC benchmarking ``Benchmarking of quantum computers'' was started in October 2023. The minimum objective is to develop a Technical Report on the topic. In this context most of the EU initiatives on QC benchmarking have been invited to present their progress, so far. Actually, several of the partners involved in these initiatives are also participating to the this CEN-CENELEC premiminary work. This work is also expected to benefit from the recently published DIN Spec ``Benchmarking quantum computers with determined KPIs'', a pre-normative document prepared in Germany. Advancing the standardization work on QC benchmarks at the EU level and building then a strong EU position on the topic appears critical in the context of starting the standardization activities on QT at the new international IEC/ISO JTC3, in a very competitive atmosphere, involving big players like the US and China. 

Precisely, in IEC/ISO JTC3, QC benchmarks have been recognized as a priority topic. Currently, a new project proposal is in preparation, covering QC benchmarks from hardware-level to application-level and beyond. It is led by France and Japan, with expected contributions from the United Kingdom, Germany, Canada, Australia and South Corea.

\section{Conclusion: Recommendations for further steps}
\label{sec:conclusion}

The document at hand outlines reasonable benchmarking protocols that cover a broad range from quantum components to quantum systems, their integration into HPC and cloud, as well as the evaluation of the performance in applications. Some of these benchmarks are currently established and spread throughout the community, other benchmarks are less established. However, very rarely these benchmarks are accompanied with a standardised implementation and evaluation software, or ideally example data (which may also facilitate continuous integration/deployment).
The \textbf{development and establishment of hardware-agnostic, statistically sound, and numerically efficient evaluation routines}
is a significant effort, on top of current activities to establish the benchmarking protocols in an open and standardised fashion.

Note that these benchmarks should not be developed in isolation but require \textbf{tight interaction and alignment with ongoing quantum activities} across the entire quantum stack: from software to hardware, from HPC to cloud. Rather than having one large activity, we are convinced that separate focused activities, addressing specific areas, e.g. components, QEC, HPC, cloud, or applications, will result in the necessary depth and width to provide standards that may be recognised on an international level.

The \textbf{integration of these benchmarks into standardisation activities} need to be covered by entities more suitable for ISO and CEN-CENELEC. Note that representatives need to be appointed on the national level, and require a significant time of coordination on a EU-wide scale to ensure that sufficient integrity of the proposed benchmarks and standards is realised to subsequently support international application.


\section*{Acknowledgments}
We gratefully acknowledge support by the European Union’s Horizon Europe Coordination and Support Action project under Grant Agreement 101070193 (QUCATS),
the research and innovation program under Grant Agreement Number 101114305 (“MILLENION-SGA1” EU Project), 
PasQuans2
OpenSuperQ+
QuTest),
the European Research Council
ERC 
(DebuQC)
the European QuantERA
(HQCC),
the European Innovation Council
(CatQubit),
The French BPI
(Usine à Chat),
the MetriQs-France program, which is itself supported by France 2030 under the French National Research Agency grant number ANR-22-QMET-0002,
the German BMBF 
(QSolid,
Hybrid++, 
QR.X, 
MuniQC-Atoms, 
DAQC, 
QuSol,
PhoQuant, 
QPIC-1),
the German BMWK 
(QuaST,
EniQmA),
the Bavarian High Tech Agenda
(Munich Quantum Valley),
and the German Research Foundation DFG
(CRC 183, FOR 2724)
for suppport.
the Austrian Research Promotion Agency under Contracts Number 896213 (ITAQC) and 897481 (HPQC). This research was funded in part by the Austrian Science Fund (FWF) [10.55776/F71]. For open access purposes, the authors have applied a CC BY public copyright license to any author accepted manuscript version arising from this submission. We further receive support from the IQI GmbH.


\printbibliography

\end{document}